\documentclass[aps,prl,twocolumn,superscriptaddress,showpacs,preprintnumbers,amsmath,amssymb]{revtex4}
\usepackage{graphicx} % Include figure files
\usepackage{dcolumn}  % Align table columns on decimal point
\usepackage{rotating}
\usepackage{threeparttable}
\usepackage[colorlinks,linkcolor=red,anchorcolor=green,citecolor=blue]{hyperref}

\newcommand{\BR}{{\cal B}}
\newcommand{\EE}{e^+e^-}

\newcommand {\tabincell}[2]{\begin{tabular}{@{}#1@{}}#2\end{tabular}}%
\begin{document}
\hyphenpenalty=10000
%\tolerance=1000
\vspace*{-3\baselineskip}
%\resizebox{!}{3cm}{\includegraphics{belle.eps}}

%\input{pub505}

%\preprint{\vbox{ \hbox{   }
%                        %\hbox{Belle DRAFT {\it 17-07}}
%                       \hbox{Intended for {\it P.R.L}}
%                       \hbox{Author: Y. B. Li, C. P. Shen, C. Z. Yuan}
%                       \hbox{Committee: J. Yelton(chair),}
%                       \hbox{ $~~~~~~~~~~~~~~~~$B. Grube, U. Tamponi }
%                        \hbox{KEK Preprint \# 2017-35}
%}
%}

\title{% \quad\\[3.0cm]
First measurements of absolute branching fractions of the $\Xi_c^0$ baryon at Belle}

%%%% >>>>> insert the authorlist here. BEFORE the abstract !!!!! <<<<<
%%%% >>>>> from the authorship confirmation web page
%%% Name the file author.tex and use \input{author} to insert into your latex file.
%\author{Author}\affiliation{affiliation}
%\collaboration{The Belle Collaboration}
%\noaffiliation
%% end author list

%%% Paper:    Xi_c0 branching fractions
%%% Journal:  Physical Review Letters
%%% Contacts: Y.B. Li (liyb@pku.edu.cn)
%%%           C.P. Shen (shencp@buaa.edu.cn)
%%%           C.Z. Yuan (yuancz@mail.ihep.ac.cn)
%%% Non-responding authors or those who said NO are commented out.
%%% ====================================================================
%%% Click the RELOAD button on your web browser to see the updated file.
%%% ====================================================================
%%% Use \input{author} to insert this material into your latex file.
%%%%% Force institutions to appear in alphabetical order when typeset.
\noaffiliation
\affiliation{University of the Basque Country UPV/EHU, 48080 Bilbao}
\affiliation{Beihang University, Beijing 100191}
%%%\affiliation{University of Bonn, 53115 Bonn}
\affiliation{Brookhaven National Laboratory, Upton, New York 11973}
\affiliation{Budker Institute of Nuclear Physics SB RAS, Novosibirsk 630090}
\affiliation{Faculty of Mathematics and Physics, Charles University, 121 16 Prague}
%%%\affiliation{Chiba University, Chiba 263-8522}
\affiliation{Chonnam National University, Kwangju 660-701}
\affiliation{University of Cincinnati, Cincinnati, Ohio 45221}
\affiliation{Deutsches Elektronen--Synchrotron, 22607 Hamburg}
%%%\affiliation{Duke University, Durham, North Carolina 27708}
\affiliation{University of Florida, Gainesville, Florida 32611}
%%%\affiliation{Department of Physics, Fu Jen Catholic University, Taipei 24205}
\affiliation{Key Laboratory of Nuclear Physics and Ion-beam Application (MOE) and Institute of Modern Physics, Fudan University, Shanghai 200443}
%%%\affiliation{Justus-Liebig-Universit\"at Gie\ss{}en, 35392 Gie\ss{}en}
\affiliation{Gifu University, Gifu 501-1193}
\affiliation{II. Physikalisches Institut, Georg-August-Universit\"at G\"ottingen, 37073 G\"ottingen}
\affiliation{SOKENDAI (The Graduate University for Advanced Studies), Hayama 240-0193}
\affiliation{Gyeongsang National University, Chinju 660-701}
\affiliation{Hanyang University, Seoul 133-791}
\affiliation{University of Hawaii, Honolulu, Hawaii 96822}
\affiliation{High Energy Accelerator Research Organization (KEK), Tsukuba 305-0801}
\affiliation{J-PARC Branch, KEK Theory Center, High Energy Accelerator Research Organization (KEK), Tsukuba 305-0801}
\affiliation{Forschungszentrum J\"{u}lich, 52425 J\"{u}lich}
%%%\affiliation{Hiroshima Institute of Technology, Hiroshima 731-5193}
\affiliation{IKERBASQUE, Basque Foundation for Science, 48013 Bilbao}
%%%\affiliation{University of Illinois at Urbana-Champaign, Urbana, Illinois 61801}
\affiliation{Indian Institute of Science Education and Research Mohali, SAS Nagar, 140306}
%%%\affiliation{Indian Institute of Technology Bhubaneswar, Satya Nagar 751007}
\affiliation{Indian Institute of Technology Guwahati, Assam 781039}
\affiliation{Indian Institute of Technology Hyderabad, Telangana 502285}
\affiliation{Indian Institute of Technology Madras, Chennai 600036}
\affiliation{Indiana University, Bloomington, Indiana 47408}
\affiliation{Institute of High Energy Physics, Chinese Academy of Sciences, Beijing 100049}
\affiliation{Institute of High Energy Physics, Vienna 1050}
%%%\affiliation{Institute for High Energy Physics, Protvino 142281}
%%%\affiliation{Institute of Mathematical Sciences, Chennai 600113}
\affiliation{INFN - Sezione di Napoli, 80126 Napoli}
\affiliation{INFN - Sezione di Torino, 10125 Torino}
\affiliation{Advanced Science Research Center, Japan Atomic Energy Agency, Naka 319-1195}
\affiliation{J. Stefan Institute, 1000 Ljubljana}
%%%\affiliation{Kanagawa University, Yokohama 221-8686}
\affiliation{Institut f\"ur Experimentelle Teilchenphysik, Karlsruher Institut f\"ur Technologie, 76131 Karlsruhe}
%%%\affiliation{Kavli Institute for the Physics and Mathematics of the Universe (WPI), University of Tokyo, Kashiwa 277-8583}
\affiliation{Kennesaw State University, Kennesaw, Georgia 30144}
\affiliation{King Abdulaziz City for Science and Technology, Riyadh 11442}
\affiliation{Department of Physics, Faculty of Science, King Abdulaziz University, Jeddah 21589}
\affiliation{Kitasato University, Sagamihara 252-0373}
\affiliation{Korea Institute of Science and Technology Information, Daejeon 305-806}
\affiliation{Korea University, Seoul 136-713}
\affiliation{Kyoto University, Kyoto 606-8502}
\affiliation{Kyungpook National University, Daegu 702-701}
\affiliation{LAL, Univ. Paris-Sud, CNRS/IN2P3, Universit\'{e} Paris-Saclay, Orsay}
\affiliation{\'Ecole Polytechnique F\'ed\'erale de Lausanne (EPFL), Lausanne 1015}
\affiliation{P.N. Lebedev Physical Institute of the Russian Academy of Sciences, Moscow 119991}
\affiliation{Faculty of Mathematics and Physics, University of Ljubljana, 1000 Ljubljana}
\affiliation{Ludwig Maximilians University, 80539 Munich}
\affiliation{Luther College, Decorah, Iowa 52101}
%%%\affiliation{Malaviya National Institute of Technology Jaipur, Jaipur 302017}
\affiliation{University of Malaya, 50603 Kuala Lumpur}
\affiliation{University of Maribor, 2000 Maribor}
\affiliation{Max-Planck-Institut f\"ur Physik, 80805 M\"unchen}
\affiliation{School of Physics, University of Melbourne, Victoria 3010}
\affiliation{University of Mississippi, University, Mississippi 38677}
\affiliation{University of Miyazaki, Miyazaki 889-2192}
\affiliation{Moscow Physical Engineering Institute, Moscow 115409}
\affiliation{Moscow Institute of Physics and Technology, Moscow Region 141700}
\affiliation{Graduate School of Science, Nagoya University, Nagoya 464-8602}
\affiliation{Kobayashi-Maskawa Institute, Nagoya University, Nagoya 464-8602}
\affiliation{Universit\`{a} di Napoli Federico II, 80055 Napoli}
%%%\affiliation{Nara University of Education, Nara 630-8528}
\affiliation{Nara Women's University, Nara 630-8506}
\affiliation{National Central University, Chung-li 32054}
\affiliation{National United University, Miao Li 36003}
\affiliation{Department of Physics, National Taiwan University, Taipei 10617}
\affiliation{H. Niewodniczanski Institute of Nuclear Physics, Krakow 31-342}
\affiliation{Nippon Dental University, Niigata 951-8580}
\affiliation{Niigata University, Niigata 950-2181}
%%%\affiliation{University of Nova Gorica, 5000 Nova Gorica}
\affiliation{Novosibirsk State University, Novosibirsk 630090}
\affiliation{Osaka City University, Osaka 558-8585}
%%%\affiliation{Osaka University, Osaka 565-0871}
\affiliation{Pacific Northwest National Laboratory, Richland, Washington 99352}
\affiliation{Panjab University, Chandigarh 160014}
\affiliation{Peking University, Beijing 100871}
\affiliation{University of Pittsburgh, Pittsburgh, Pennsylvania 15260}
%%%\affiliation{Punjab Agricultural University, Ludhiana 141004}
%%%\affiliation{Research Center for Electron Photon Science, Tohoku University, Sendai 980-8578}
%%%\affiliation{Research Center for Nuclear Physics, Osaka University, Osaka 567-0047}
\affiliation{Theoretical Research Division, Nishina Center, RIKEN, Saitama 351-0198}
%%%\affiliation{RIKEN BNL Research Center, Upton, New York 11973}
%%%\affiliation{Saga University, Saga 840-8502}
\affiliation{University of Science and Technology of China, Hefei 230026}
\affiliation{Seoul National University, Seoul 151-742}
%%%\affiliation{Shinshu University, Nagano 390-8621}
\affiliation{Showa Pharmaceutical University, Tokyo 194-8543}
\affiliation{Soongsil University, Seoul 156-743}
%%%\affiliation{University of South Carolina, Columbia, South Carolina 29208}
\affiliation{Stefan Meyer Institute for Subatomic Physics, Vienna 1090}
\affiliation{Sungkyunkwan University, Suwon 440-746}
\affiliation{School of Physics, University of Sydney, New South Wales 2006}
\affiliation{Department of Physics, Faculty of Science, University of Tabuk, Tabuk 71451}
\affiliation{Tata Institute of Fundamental Research, Mumbai 400005}
%%%\affiliation{Excellence Cluster Universe, Technische Universit\"at M\"unchen, 85748 Garching}
\affiliation{Department of Physics, Technische Universit\"at M\"unchen, 85748 Garching}
%%%\affiliation{Toho University, Funabashi 274-8510}
%%%\affiliation{Tohoku Gakuin University, Tagajo 985-8537}
\affiliation{Department of Physics, Tohoku University, Sendai 980-8578}
\affiliation{Earthquake Research Institute, University of Tokyo, Tokyo 113-0032}
\affiliation{Department of Physics, University of Tokyo, Tokyo 113-0033}
\affiliation{Tokyo Institute of Technology, Tokyo 152-8550}
\affiliation{Tokyo Metropolitan University, Tokyo 192-0397}
%%%\affiliation{Tokyo University of Agriculture and Technology, Tokyo 184-8588}
%%%\affiliation{Utkal University, Bhubaneswar 751004}
\affiliation{Virginia Polytechnic Institute and State University, Blacksburg, Virginia 24061}
\affiliation{Wayne State University, Detroit, Michigan 48202}
\affiliation{Yamagata University, Yamagata 990-8560}
\affiliation{Yonsei University, Seoul 120-749}
\author{Y.~B.~Li}\affiliation{Peking University, Beijing 100871} % Peking
\author{C.~P.~Shen}\affiliation{Beihang University, Beijing 100191} % Beihang
\author{C.~Z.~Yuan}\affiliation{Institute of High Energy Physics, Chinese Academy of Sciences, Beijing 100049} % IHEP
% \author{A.~Abdesselam}\affiliation{Department of Physics, Faculty of Science, University of Tabuk, Tabuk 71451} % Tabuk
\author{I.~Adachi}\affiliation{High Energy Accelerator Research Organization (KEK), Tsukuba 305-0801}\affiliation{SOKENDAI (The Graduate University for Advanced Studies), Hayama 240-0193} % KEK
% \author{K.~Adamczyk}\affiliation{H. Niewodniczanski Institute of Nuclear Physics, Krakow 31-342} % Krakow
% \author{J.~K.~Ahn}\affiliation{Korea University, Seoul 136-713} % Korea
\author{H.~Aihara}\affiliation{Department of Physics, University of Tokyo, Tokyo 113-0033} % Tokyo
\author{S.~Al~Said}\affiliation{Department of Physics, Faculty of Science, University of Tabuk, Tabuk 71451}\affiliation{Department of Physics, Faculty of Science, King Abdulaziz University, Jeddah 21589} % Tabuk
% \author{K.~Arinstein}\affiliation{Budker Institute of Nuclear Physics SB RAS, Novosibirsk 630090}\affiliation{Novosibirsk State University, Novosibirsk 630090} % BINP
% \author{Y.~Arita}\affiliation{Graduate School of Science, Nagoya University, Nagoya 464-8602} % Nagoya
\author{D.~M.~Asner}\affiliation{Brookhaven National Laboratory, Upton, New York 11973} % BNL
% \author{H.~Atmacan}\affiliation{University of South Carolina, Columbia, South Carolina 29208} % SouthCarolina
% \author{V.~Aulchenko}\affiliation{Budker Institute of Nuclear Physics SB RAS, Novosibirsk 630090}\affiliation{Novosibirsk State University, Novosibirsk 630090} % BINP
\author{T.~Aushev}\affiliation{Moscow Institute of Physics and Technology, Moscow Region 141700} % MIPT
\author{R.~Ayad}\affiliation{Department of Physics, Faculty of Science, University of Tabuk, Tabuk 71451} % Tabuk
% \author{T.~Aziz}\affiliation{Tata Institute of Fundamental Research, Mumbai 400005} % Tata
% \author{V.~Babu}\affiliation{Tata Institute of Fundamental Research, Mumbai 400005} % Tata
\author{I.~Badhrees}\affiliation{Department of Physics, Faculty of Science, University of Tabuk, Tabuk 71451}\affiliation{King Abdulaziz City for Science and Technology, Riyadh 11442} % Tabuk
% \author{S.~Bahinipati}\affiliation{Indian Institute of Technology Bhubaneswar, Satya Nagar 751007} % IITB
% \author{A.~M.~Bakich}\affiliation{School of Physics, University of Sydney, New South Wales 2006} % Sydney
% ...
% Lebedev
% \author{V.~Zhulanov}\affiliation{Budker Institute of Nuclear Physics SB RAS, Novosibirsk 630090}\affiliation{Novosibirsk State University, Novosibirsk 630090} % BINP
% \author{T.~Zivko}\affiliation{J. Stefan Institute, 1000 Ljubljana} % Ljubljana
% \author{A.~Zupanc}\affiliation{Faculty of Mathematics and Physics, University of Ljubljana, 1000 Ljubljana}\affiliation{J. Stefan Institute, 1000 Ljubljana} % Ljubljana
% \author{N.~Zwahlen}\affiliation{\'Ecole Polytechnique F\'ed\'erale de Lausanne (EPFL), Lausanne 1015} % Lausanne
\collaboration{The Belle Collaboration}

\begin{abstract}

We present the first measurements of absolute branching fractions of $\Xi_c^0$ decays into $\Xi^- \pi^+$, $\Lambda K^- \pi^+$, and $p K^- K^- \pi^+$ final states. The measurements are made using a data set comprising  $(772\pm 11)\times 10^{6}$ $B\bar{B}$ pairs collected at the $\Upsilon(4S)$ resonance with the Belle detector at the KEKB $e^+e^-$ collider. We first measure the absolute branching fraction for $B^- \to \bar{\Lambda}_c^- \Xi_c^0$ using a missing-mass technique; the result is $\BR(B^- \to \bar{\Lambda}_c^- \Xi_c^0) = (9.51 \pm 2.10 \pm 0.88) \times 10^{-4}$.
We subsequently measure the product branching fractions
$\BR(B^- \to \bar{\Lambda}_c^- \Xi_c^0)\BR(\Xi_c^0 \to \Xi^- \pi^+)$, $\BR( B^- \to \bar{\Lambda}_c^- \Xi_c^0) \BR(\Xi_c^0 \to \Lambda K^- \pi^+)$, and $\BR( B^- \to \bar{\Lambda}_c^- \Xi_c^0) \BR(\Xi_c^0 \to p K^- K^- \pi^+)$ with improved precision.
Dividing these product branching fractions by the result for $B^- \to \bar{\Lambda}_c^- \Xi_c^0$ yields the following branching fractions:
$\BR(\Xi_c^0 \to \Xi^- \pi^+)= (1.80 \pm 0.50 \pm 0.14)\%$, $\BR(\Xi_c^0 \to \Lambda K^- \pi^+)=(1.17 \pm 0.37 \pm 0.09)\%$, and $\BR(\Xi_c^0 \to p K^- K^- \pi^+)=(0.58 \pm 0.23 \pm 0.05)\%.$ For the above branching fractions,  the first uncertainties are statistical and the second are systematic.
Our result for $\BR(\Xi_c^0 \to \Xi^- \pi^+)$ can be combined with $\Xi_c^0$ branching fractions measured
relative to $\Xi_c^0 \to \Xi^- \pi^+$ to yield other absolute $\Xi_c^0$ branching fractions.
\end{abstract}

 %13.30.-a,13.25.-k,}

\maketitle

%%%% >>>> keep the final version single-spaced
\tighten

{\renewcommand{\thefootnote}{\fnsymbol{footnote}}}
\setcounter{footnote}{0}
%\linenumbers

%%%%%%%%%%%%%%%%%%%%%%%%%%%%%%%%%%%%%%
%%%%%%%%%   introduction  %%%%%%%%%%%%
%%%%%%%%%%%%%%%%%%%%%%%%%%%%%%%%%%%%%%

Half a century after the theory of Quantum Chromodynamics (QCD) was developed,
understanding the non-perturbative property of the strong interaction still remains a challenge.
%Understanding the non-perturbative property of the strong interaction is still challenging nowadays with half a century passed since the Quantum Chromodynamics (QCD) was founded.
Weak decays of charmed hadrons play a unique role in the study of strong interactions, as the charm mass scale is near the boundary between perturbative and non-perturbative QCD. The charmed-baryon sector offers an excellent laboratory for testing heavy-quark symmetry and light-quark chiral symmetry, both of which have important implications for the low-energy dynamics of heavy baryons interacting with Goldstone bosons~\cite{gold}.
In exclusive charm decays, the heavy-quark expansion does not work, and experimental data is needed to extract non-perturbative quantities in the decay amplitudes~\cite{input1, input2, input3, input4}. Decays of charmed baryons with an additional quark and spin of 1/2
provide complementary information to that of charm-meson decays.

Unlike in the charmed-meson sector, where $D^{0}$, $D^+$, and $D_s^+$ decays are all well-measured,
in the charm baryon sector only $\Lambda_c^+$ absolute branching fractions have been measured~\cite{lc_belle,lc_bes}.
Thus, the branching fractions of $\Xi_c^0$ baryons are all measured relative to
the $\Xi_c^0 \to \Xi^-\pi^+$ mode. Thus a measurement of
%Absolute branching fractions of the other two particles of the antitriplet of charmed baryons, $\Xi_c^{0,+}$, are still unmeasured.
%So far, the branching fractions of $\Xi_c^0$ decays are all  measured relative to the $\Xi^-\pi^+$ mode.
the absolute branching fraction $\BR(\Xi_c^0 \to \Xi^- \pi^+)$
is needed to determine the absolute branching fractions of other $\Xi_c^0$ decays.
In charmed-baryon decays, non-factorizable contributions to the decay amplitude are important, and a variety of models have been developed to predict
the decay rate in such processes~\cite{QCD-theory1,QCD-theory2,QCD-theory3,QCD-theory4,QCD-theory5,QCD-theory6,QCD-theory7,QCD-theory8,Xc-theory2,Xc-theory3}.
For example, the $\BR(\Xi_c^0 \to \Xi^- \pi^+)$ has been predicted to be
0.74\% or 1.12\%~\cite{QCD-theory8},
$(2.24 \pm 0.34)\%$~\cite{Xc-theory2}, and $(1.91\pm0.17)\%$~\cite{Xc-theory3}.
Experimental information is crucial to validate these models as well as to constrain the model parameters.

The $\BR(\Xi_c^0 \to \Lambda K^- \pi^+)$ and $\BR(\Xi_c^0 \to p K^- K^- \pi^+)$ have been
measured relative to $\BR(\Xi_c^0 \to \Xi^- \pi^+)$ to be
$1.07 \pm 0.12 \pm 0.07$ and $0.33 \pm 0.03\pm 0.03 $~\cite{belle-old2},
respectively. The decay $\Xi_c^0 \to p K^- K^- \pi^+$ plays a key role in many bottom-baryon studies at LHCb~\cite{LHCB1,LHCB2}. The decay $B^- \to \bar{\Lambda}_c^- \Xi_c^0$, which proceeds via a $b \to c \bar{c} s$ transition, has a branching fraction predicted to be of the order $10^{-3}$~\cite{BXL_theroy}.
However, this has not been measured because the absolute branching fractions of $\Xi_c^0$ are unknown.  The measured product branching fractions are $\BR(B^- \to \bar{\Lambda}_c^- \Xi_c^0) \BR(\Xi_c^0 \to \Xi^- \pi^+) = (2.4 \pm 0.9) \times 10 ^{-5}$ and $\BR(B^- \to \bar{\Lambda}_c^- \Xi_c^0) \BR(\Xi_c^0 \to \Lambda K^- \pi^+) = (2.1 \pm 0.9) \times 10 ^{-5}$~\cite{PDG,belle-old1, babar-old2}.

%In this Letter, we perform an analysis of $B^- \to \bar{\Lambda}_c^- \Xi_c^0$ with ${\bar{\Lambda}}_c^-$
%reconstructed via its decays into $\bar{p} K^+ \pi^-$ and $\bar{p} K_{S}^{0}$, and $\Xi_c^0$ reconstructed either inclusively
%or exclusively decay modes into
%$\Xi^- \pi^+$, $\Lambda K^- \pi^+$, and $p K^- K^- \pi^+$~\cite{charge-conjugate},
%and measure the absolute branching fractions of these $\Xi_c^0$ decays for the first time.
%\textcolor[rgb]{0.00,0.07,1.00}{For the inclusively reconstructed $\Xi_c^0$ decays}, the number of $\Xi_{c}^{0}$ baryons is determined by
%reconstructing the recoiling $\bar{\Lambda}_{c}^{-}$ baryons in the events of
%$B^{-} \to \bar{\Lambda}_{c}^{-} \Xi_{c}^{0}$ with hadronic tagging and the
%$\BR(B^- \to \bar{\Lambda}_c^- \Xi_c^0)$ is measured.
%For exclusive $\Xi_c^0$ decays, all the final states are reconstructed without tagging,
%and the product branching fractions of $B^{-} \to \bar{\Lambda}_{c}^{-} \Xi_{c}^{0}$ and $\Xi_c^0$ decays
%are measured. The absolute $\BR(\Xi_c^0 \to \Xi^- \pi^+)$ can then be determined using
%$\BR(\Xi_c^0 \to \Xi^-
%\pi^+) \equiv \BR(B^-\to \bar{\Lambda}_c^- \Xi_c^0) \BR(\Xi_c^0 \to \Xi^- \pi^+)/\BR(B^- \to \bar{\Lambda}_c^- \Xi_c^0)$
%for example. This analysis is based on the full data sample collected at the $\Upsilon(4S)$ resonance by the Belle detector~\cite{Belle} at
%the KEKB collider~\cite{KEKB}.

In this Letter, we perform an analysis of $B^- \to \bar{\Lambda}_c^- \Xi_c^0$ with ${\bar{\Lambda}}_c^-$
reconstructed via $\bar{p} K^+ \pi^-$ and $\bar{p} K_{S}^{0}$ modes, and $\Xi_c^0$ reconstructed
both inclusively and exclusively via $\Xi^- \pi^+$, $\Lambda K^- \pi^+$, and
$p K^- K^- \pi^+$ modes~\cite{charge-conjugate}.
We present first a measurement of the absolute branching fraction for
$B^- \to \bar{\Lambda}_c^- \Xi_c^0$ using a missing-mass technique. For this analysis we fully reconstruct the
tag-side $B^+$ decay. We subsequently measure the product branching fractions
$\BR(B^- \to \bar{\Lambda}_c^- \Xi_c^0)\BR(\Xi_c^0 \to \Xi^- \pi^+)$, $\BR( B^- \to \bar{\Lambda}_c^- \Xi_c^0) \BR(\Xi_c^0 \to \Lambda K^- \pi^+)$, and $\BR( B^- \to \bar{\Lambda}_c^- \Xi_c^0) \BR(\Xi_c^0 \to p K^- K^- \pi^+)$. For these measurements we do not reconstruct the
recoiling $B^+$ decay, as the signal decays are fully reconstructed.
Dividing these product branching fractions by $\BR(B^- \to \bar{\Lambda}_c^- \Xi_c^0)$ yields the branching fractions
$\BR(\Xi_c^0 \to \Xi^- \pi^+)$, $\BR(\Xi_c^0 \to \Lambda K^- \pi^+)$, and $\BR(\Xi_c^0 \to p K^- K^- \pi^+)$.

This analysis is based on the full data sample of 702.6 fb$^{-1}$
collected at the $\Upsilon(4S)$ resonance by the Belle detector~\cite{Belle} at
the KEKB asymmetric-energy $e^{+}e^{-}$ collider~\cite{KEKB}.
%The Belle detector is a large-solid-angle magnetic spectrometer that consists of a silicon vertex detector, a
%50-layer central drift chamber (CDC), an array of aerogel threshold  Cherenkov  counters  (ACC),  a barrel-like
%arrangement of time-of-flight (TOF) scintillation counters, and an electromagnetic calorimeter composed of CsI(Tl)
%crystals located inside a superconducting solenoid coil that provides a 1.5 T magnetic field. An iron
%flux-return yoke located outside of the coil is instrumented to detect $K_L^0$ mesons and to identify muons.
The detector is described in detail elsewhere~\cite{Belle}.

To optimize signal selection criteria and calculate the signal reconstruction efficiency,
we use Monte Carlo (MC) simulated events. Signal events of $B$ meson decays are generated
using {\sc evtgen}~\cite{evtgen}, while
inclusive $\Xi_c^0$ decays are generated using {\sc pythia}~\cite{pythia}.
The MC events are processed with a detector simulation based on {\sc
geant3}~\cite{geant3}. MC samples of
$\Upsilon(4S)\to B \bar{B}$ events with $B=B^+$ or $B^0$, and $e^+e^- \to q
\bar{q}$ events with $q=u,~d,~s,~c$ at $\sqrt{s}=10.58$ GeV are used
as background samples.

To select signal candidates, well-reconstructed tracks and particle identification
are performed using the same method as in Ref.~\cite{liyb},
as well as the $\Lambda \to p \pi^-$ and $K_{S}^{0} \to \pi^+ \pi^-$ candidates~\cite{liyb}.

%The $K_{S}^{0}$ candidates are reconstructed from pairs of
%oppositely charged tracks, treated as pions, and identified by a
%multivariate analysis using a Neurobayes neural network~\cite{NN} based on two
%sets of input variables~\cite{NN-input}.
%The $\Lambda$ candidates are reconstructed from the decay $\Lambda \to p \pi^-$ and
%selected if the $p \pi^-$ invariant mass is within 5~MeV/$c^2$
%of the $\Lambda$ nominal mass~\cite{PDG}, which corresponds to 3 times the resolution ($\sigma$).

For the inclusive analysis of the $\Xi_c^0$ decay,
the tag-side $B^+$ meson candidate, $B^+_{\rm tag}$,
is reconstructed using a neural network based on a
full hadron-reconstruction algorithm~\cite{FR}. Each $B^+_{\rm tag}$
candidate has an associated output value ${O}_{\rm NN}$ from the multivariate analysis that
ranges from 0 to 1.
A candidate with larger ${O}_{\rm NN}$ is more likely to be a true $B$ meson.
If multiple $B_{\rm tag}^+$ candidates are found in an event, the
candidate with the largest ${O}_{\rm NN}$ is selected.
To improve the purity of the $B^+_{\rm tag}$ sample, we require
${O}_{\rm NN} > 0.005$, $M_{\rm bc}^{\rm tag} > 5.27$~GeV/$c^{2}$,
and $|\Delta E^{\rm tag}|<0.04$~GeV, where the latter two intervals
correspond to approximately $3\sigma$ in resolution.
The variables $M_{\rm bc}^{\rm tag}$ and $\Delta E^{\rm tag}$ are defined as
$M_{\rm bc}^{\rm tag} \equiv
\sqrt{E_{\rm beam}^{2} - |\sum_i \overrightarrow{p}^{\rm tag}_{i}|^2}$
and $\Delta E^{\rm tag} \equiv \sum_i E^{\rm tag}_{i}
- E_{\rm beam}$,
where $E_{\rm beam}\equiv \sqrt{s}/2$ is the beam energy, and ($E^{\rm tag}_{i}$, $\overrightarrow{p}^{\rm tag}_{i}$)
is the four-momentum of the $B_{\rm tag}^+$ daughter $i$ in the $\EE$
center-of-mass system (CMS).
After reconstructing a $B^+_{\rm tag}$ candidate, ${\bar{\Lambda}_c^-} \to \bar{p} K^{+} \pi^-$ and
${\bar{\Lambda}_c^-} \to \bar{p} K_{S}^{0}$ decays are reconstructed from among the remaining tracks.
We perform a fit for the decay vertex and require that $\chi^{2}_{\rm vertex}/{\rm n.d.f.} < 15$,
where ${\rm n.d.f.}$ is the number of degrees of freedom. If there
is more than one ${\bar{\Lambda}_c^-}$ candidate in an event, the candidate with
the smallest $\chi^{2}_{\rm vertex}/{\rm n.d.f.}$ is selected.
%The percentage of events with multiple candidates is 3.5\%.
We define a $\bar{\Lambda}_c^-$ signal region $|M^{}_{\bar{p} K^+ \pi^-/\bar{p} K_{S}^{0}} - m_{\bar{\Lambda}_c^-}| < 10$~MeV/$c^{2}$ (3.0$\sigma$),
where
%$M^{}_{\bar{p} K^+ \pi^-/\bar{p} K_{S}^{0}}$  is the $\bar{p} K^+ \pi^-$/$\bar{p} K_{S}^{0}$ invariant mass and
$m_{\bar{\Lambda}_c^-}$ is the nominal mass of the $\bar{\Lambda}_c^-$~\cite{PDG}.
% and the mass range corresponds to $3\sigma$.

%Among the remaining tracks, ${\bar{\Lambda}_c^-}$ candidates are selected and
%vertex fits are performed. We require $\chi^{2}_{\rm vertex}/{\rm n.d.f.} < 15$, where
%${\rm n.d.f.}$ is the number of degrees of freedom of the vertex fit. If there
%is more than one ${\bar{\Lambda}_c^-}$ candidate in an event, the candidate with
%the smallest $\chi^{2}_{\rm vertex~fit}/{\rm n.d.f.}$ is selected (the percentage of events with multiple candidates is 3.5\%).
%A $3 \sigma$ $\bar{\Lambda}_c^-$ signal region is defined by
%$|M_{\bar{\Lambda}_c^-} - m_{\bar{\Lambda}_c^-}| < 10$~MeV/$c^{2}$,
%where $M_{\bar{\Lambda}_c^-}$ is the invariant mass of $\bar{p} K^{+} \pi^-$ or $\bar{p} K_{S}^{0}$,
%and $m_{\bar{\Lambda}_c^-}$ is the nominal mass of the
%$\bar{\Lambda}_c^-$ baryon~\cite{PDG}.

\begin{figure}[htbp]
    \begin{center}
        \includegraphics[width=6cm]{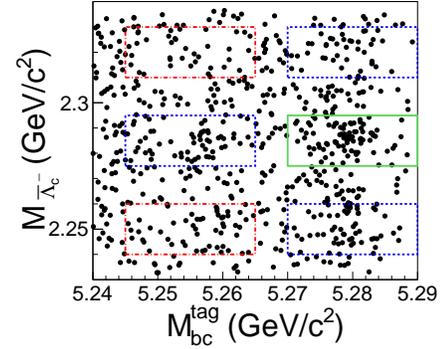}
\caption{The distribution of $M_{\rm bc}^{\rm tag}$ of $B_{\rm
tag}^+$ versus $M_{\bar{\Lambda}_c^-}$ of selected $B^- \to \bar{\Lambda}_c^- \Xi_c^0$ candidates with $\Xi_c^0 \to \rm anything$, summed over the two reconstructed $\bar{\Lambda}_c^-$ decay modes.
The solid box shows the signal region, and the dashed and dash-dotted boxes define the
$M_{\rm bc}^{\rm tag}$ and $M_{\bar{\Lambda}_c^-}$ sidebands described in the
text.}\label{FR_mbc_mlc_com}
    \end{center}
\end{figure}

\begin{figure}[htbp]
    \begin{center}
        \includegraphics[width=7cm]{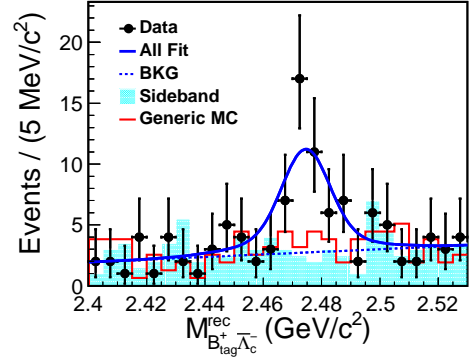}
\caption{The fit to the $M_{B_{\rm tag}^+\bar{\Lambda}_c^-}^{\rm rec}$
distribution of the selected candidate events. The points with error bars represent the data, the solid blue curve is the best fit, the dashed
curve is the fitted background, the cyan shaded histogram is from the
scaled $M_{\rm bc}^{\rm tag}$ and $M_{\bar{\Lambda}_c^-}$
sidebands, the red open histogram is from the sum of the
MC-simulated contributions from the $\EE \to q\bar{q}$ with
$q=u,~d,~s,~c$, and $\Upsilon(4S) \to B \bar{B}$ generic-decay
backgrounds with the number of events normalized to the number of events
from the normalized $M^{\rm tag}_{\rm bc}$ and
$M_{\bar{\Lambda}_c^-}$ sidebands.}\label{FR_xic_com}
    \end{center}
\end{figure}

The `recoil mass' of the daughter $X$ in $B^- \to \bar{\Lambda}_c^- + X$ is calculated
using $M^{\rm recoil}_{B_{\rm tag}^{+}\bar{\Lambda}_{c}^{-}} = \sqrt{(P_{\rm CMS} - P_{B^{+}_{\rm tag}} - P_{\bar{\Lambda}_{c}^{-}})^{2}}$,
where $P_{\rm CMS},~P_{B_{\rm tag}^{+}}$, and $P_{\bar{\Lambda}_{c}^{-}}$ are the four momenta of the
initial $\EE$ system, the tagged $B^+$ meson, and the reconstructed $\bar{\Lambda}_{c}^{-}$ baryon.
To improve the recoil mass resolution, we use $M_{B_{\rm tag}^{+}\bar{\Lambda}_{c}^{-}}^{\rm rec}\equiv M^{\rm recoil}_{B_{\rm tag}^{+}\bar{\Lambda}_{c}^{-}} + M_{B_{\rm tag}^{+}} - m_{B} + M_{\bar{\Lambda}_{c}^{-}} -m_{\bar{\Lambda}_{c}^{-}}$,
where $M_{B_{\rm tag}^{+}}$  is the invariant mass of the $B_{\rm tag}^{+}$ candidate,
$M_{\bar{\Lambda}_{c}^{-}}$ is the reconstructed mass of the $\bar{\Lambda}_c^-$ candidate, and
 $m_{B}$ is the nominal mass of the $B$ meson~\cite{PDG}.
The distribution of $M_{\rm bc}^{\rm tag}$ of the $B_{\rm tag}^+$ candidates versus $M_{\bar{\Lambda}_c^-}$
of the selected $B^- \to \bar{\Lambda}_c^- \Xi_c^0$ signal candidates summed over the two reconstructed
$\bar{\Lambda}_c^-$ decay modes is shown in Fig.~\ref{FR_mbc_mlc_com},
for $2.40<M_{B_{\rm tag}^{+}\bar{\Lambda}_{c}^{-}}^{\rm rec}<2.53$ GeV/$c^2$.
We observe a significant excess of $B^- \to \bar{\Lambda}_c^- \Xi_c^0$ candidates in the
signal region denoted as the solid box in Fig.~\ref{FR_mbc_mlc_com}.
To check for possible peaking backgrounds, we define $M_{\rm bc}^{\rm tag}$ and $M_{\bar{\Lambda}_c^-}$ sidebands,
represented by the dashed and dash-dotted boxes in Fig.~\ref{FR_mbc_mlc_com}.
Each sideband box is the same size as the signal box.
The background contribution in the signal box is estimated
using half the number of events in the blue dashed sideband boxes minus one fourth the number of events in
the red dash-dotted sideband boxes.
The $M_{B_{\rm tag}^{+}\bar{\Lambda}_{c}^{-}}^{\rm rec}$ distribution of events
in both the signal and sideband boxes is shown in Fig.~\ref{FR_xic_com}.
No peaking backgrounds
in the studied recoil $\Xi_c^0$ mass region are found in the
$M_{\rm bc}^{\rm tag}$ and $M_{\bar{\Lambda}_c^-}$ sideband events, as shown with the
shaded histogram in Fig.~\ref{FR_xic_com}.

%The selection of
%sideband regions is determined from a two-dimensional (2D)
%unbinned maximum likelihood fit to $M^{\rm tag}_{\rm bc}$ and
%$M_{\bar{\Lambda}_c^-}$ distributions so that the normalized
%contribution from $M^{\rm tag}_{\rm bc}$ and
%$M_{\bar{\Lambda}_c^-}$ sidebands is the half of events in blue
%dashed boxes minus half of events in red dashed boxes.

To extract the $\Xi_c^0$ signal yield, an
unbinned maximum-likelihood fit is performed to the $M_{B_{\rm
tag}^+\bar{\Lambda}_c^-}^{\rm rec}$ distribution. A
double-Gaussian function (its parameters are fixed to those from a fit to the MC-simulated signal
distribution) is used to model the $\Xi_c^0$ signal shape,
and a first-order polynomial is taken as the background shape.
%Due to the limited size of the data sample, the parameters of the double-Gaussian
%function are fixed to those from a fit to the MC-simulated signal
%distribution.
The fit results are shown in Fig.~\ref{FR_xic_com}.

The fitted $\Xi_c^0$ signal yield is $N_{\Xi_c^0} = 40.9 \pm 9.0$,
with a statistical significance of $5.5\sigma$. The significance is calculated using
$\sqrt{-2\ln(\mathcal{L}_{0}/\mathcal{L}_{\rm max})}$, where $\mathcal{L}_{0}$ and $\mathcal{L}_{\rm max}$ are the likelihoods of the fits without and with a signal component, respectively.
The $\BR(B^- \to \bar{\Lambda}_c^- \Xi_c^0)$ is calculated using
$N_{\Xi_c^0}/[N_{B^-}(\varepsilon_{1} \BR_1 + \varepsilon_{2}
\BR_2)]$. In this expression, $\BR_1=\BR(\bar{\Lambda}_c^- \to \bar{p} K^+ \pi^-)$,
$\BR_2=\BR(\bar{\Lambda}_c^- \to \bar{p} K_S^0 )\BR(K_S^0 \to
\pi^+\pi^-)$, and $N_{B^-}=2N_{\Upsilon(4S)}\BR(\Upsilon(4S)\to B^+B^-)$,
where $N_{\Upsilon(4S)}$ is the number of $\Upsilon(4S)$ events, and the
$\BR[\Upsilon(4S)\to B^+B^-]=(51.4\pm0.6)\%$~\cite{PDG}. The reconstruction
efficiencies $\varepsilon_{1}$ and $\varepsilon_{2}$
of the two $\bar{\Lambda}_c^-$ decay modes are obtained from MC simulation.
The $\BR(\bar{\Lambda}_c^- \to \bar{p} K^+
\pi^-)$, $\BR(\bar{\Lambda}_c^- \to \bar{p} K_S^0 )$, and
$\BR(K_S^0 \to \pi^+\pi^-)$ are taken from Ref.~\cite{PDG}.
The result is $\BR(B^- \to \bar{\Lambda}_c^- \Xi_c^0)=[9.51 \pm 2.10(\rm stat.)] \times 10^{-4}$.

%The branching fraction $\BR(B^- \to \bar{\Lambda}_c^- \Xi_c^0)$
%is calculated using $\frac{N_{\Xi_c^0}} {2 N_{B^-} [\varepsilon_{1}
%\BR(\bar{\Lambda}_c^- \to \bar{p} K^+ \pi^-) + \varepsilon_{2}
%\BR(\bar{\Lambda}_c^- \to \bar{p} K_S^0 ) \BR(K_S^0 \to
%\pi^+\pi^-) ] }$ and the result is $[9.51 \pm 2.10(\rm stat.)] \times 10^{-4}$,
%where $N_{B^-}=N_{\Upsilon(4S)}\BR(\Upsilon(4S)\to B^+B^-)$
%($N_{\Upsilon(4S)}$ is the number of $\Upsilon(4S)$ events and
%$\BR(\Upsilon(4S)\to B^+B^-)=(51.4\pm0.6)\%$~\cite{PDG}).
%The reconstruction efficiencies $\varepsilon_{1}$ and $\varepsilon_{2}$
%of the two $\bar{\Lambda}_c^-$ decay modes are obtained from MC simulation.
%The branching fractions $\BR(\bar{\Lambda}_c^- \to \bar{p} K^+
%\pi^-)$, $\BR(\bar{\Lambda}_c^- \to \bar{p} K_S^0 )$, and
%$\BR(K_S^0 \to \pi^+\pi^-)$ are taken from Ref.~\cite{PDG}.

%%%%%%%%%%%%%%%%%%%%%%%%%%%%%%%%%%%%%
%%%%%%%   exclusive     %%%%%%%%%%%%%
%%%%%%%%%%%%%%%%%%%%%%%%%%%%%%%%%%%%%

For the analysis of the exclusive $\Xi_c^0$ decays, we
again use $B^- \to \bar{\Lambda}_c^- \Xi_c^0$ decays in which
$\bar{\Lambda}_c^-  \to (\bar{p} K^+ \pi^-,~\bar{p} K_S^0)$.
However, instead of reconstructing the tag-side $B^+_{\rm tag}$,
we fully reconstruct the $\Xi_c^0$ decay in the final states
$\Xi^- \pi^+$, $\Lambda K^- \pi^+$, and $p K^- K^-
\pi^+$, where $\Xi^-\to \Lambda \pi^-$
and $\Lambda \to p \pi^-$. Fits to the $B^-$, $\Xi_c^0$, and
$\Xi^-$ decay vertices are performed.
%For the analysis of the exclusive $\Xi_c^0$ decays,
%instead of tagging a $B$ meson candidate, we reconstruct
%$\Xi_c^0$ from $\Xi^- \pi^+$, $\Lambda K^- \pi^+$, and $p K^- K^-
%\pi^+$ final states with $\Xi^-\to \Lambda \pi^-$
%and $\Lambda \to p \pi^-$, separately. Vertex fits to the $B^-$, $\Xi_c^0$, and
%$\Xi^-$ candidates are performed.
If there is more than one $B^-$ candidate in an event,
the one with the smallest $\chi^{2}_{\rm vertex}/{\rm n.d.f.}$ from the $B^-$ vertex fit is selected.
%The percentage of events with multiple candidates is 3.8\%.
We subsequently require $\chi^{2}_{\rm vertex}/{\rm n.d.f.} <50,~15$, and 15 for
reconstructed $B^-$, $\Xi_c^0$, and $\Xi^-$ candidates, respectively.
%the corresponding selection efficiencies are 96\%, 95\%, and 95\%.
The $\Xi^-$ and $\Xi_c^0$ signal ranges are
defined as $|M^{}_{\Lambda\pi^-}  - m_{\Xi^-}| < 10$~MeV/$c^2$ and
$|M_{\Xi_c^0} - m_{\Xi_c^0}| < 20$~MeV/$c^2$  ($3.0\sigma$), where
$M^{}_{\Lambda\pi^-}$ and $M_{\Xi_c^0}$ are the invariant masses of the selected
$\Xi^-$ and $\Xi_c^0$ candidates, and
$m_{\Xi^-}$ and $m_{\Xi_c^0}$ are
the nominal masses of $\Xi^-$ and $\Xi_c^0$~\cite{PDG}. The $\bar{\Lambda}_c^-$
signal interval is the same as
in the inclusive analysis of $\Xi_c^0$ decays.
%As the continuum background level is low,
%further continuum suppression is not necessary.
The $B^-$ signal candidates are identified using the beam-energy-constrained
mass $M_{\rm bc}$ and the energy difference $\Delta E$,
where $M_{\rm bc}$ and $\Delta E$ are calculated in the same manner as done for $B^+_{\rm tag}$ candidates, but, here,
tracks from the $B^-$ signal candidate decay are used. % instead of tracks from the tag side.

\begin{figure*}[htbp]
    \begin{center}
        \includegraphics[width=5cm,height=3.65cm]{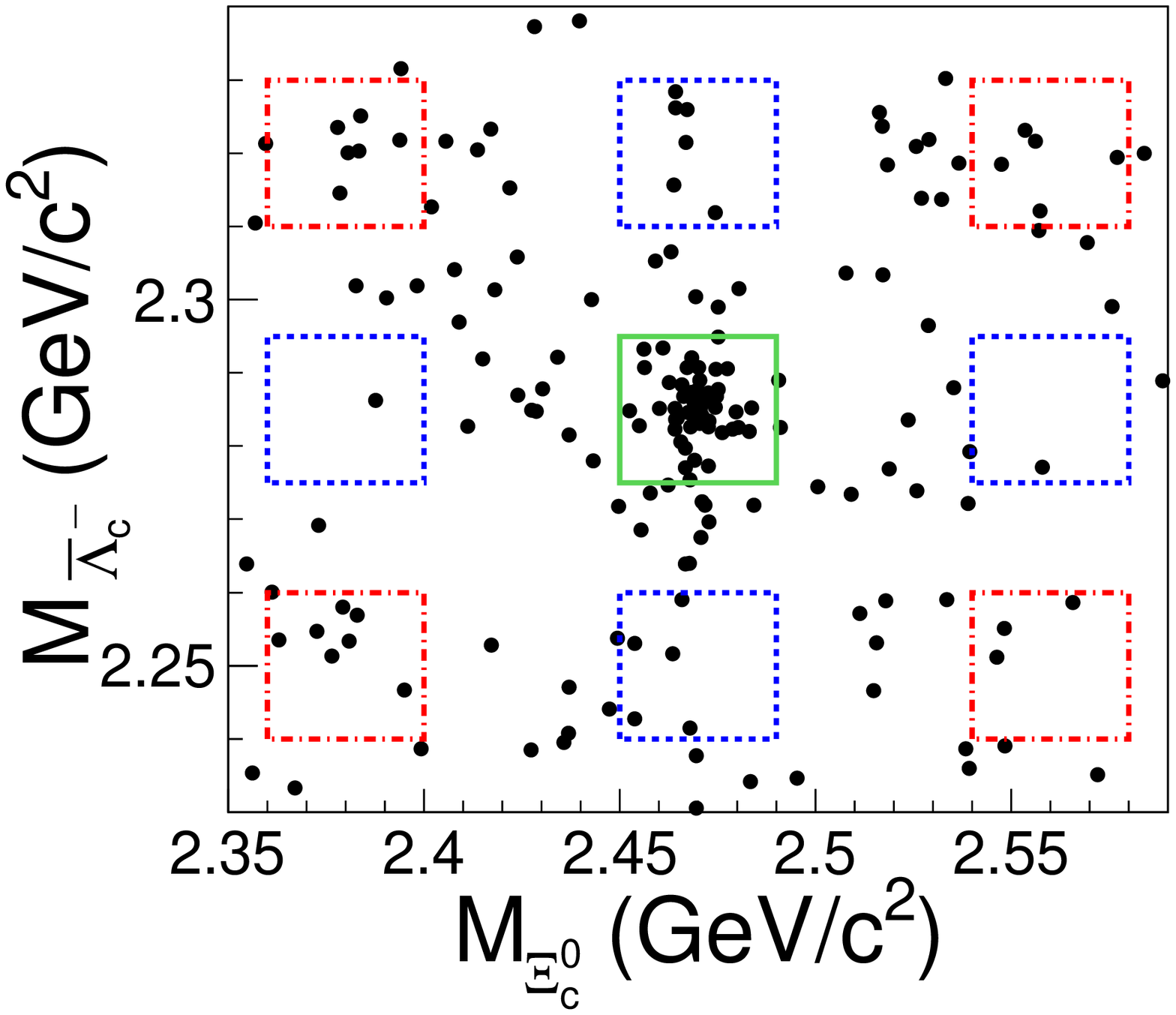}
        \includegraphics[width=5cm]{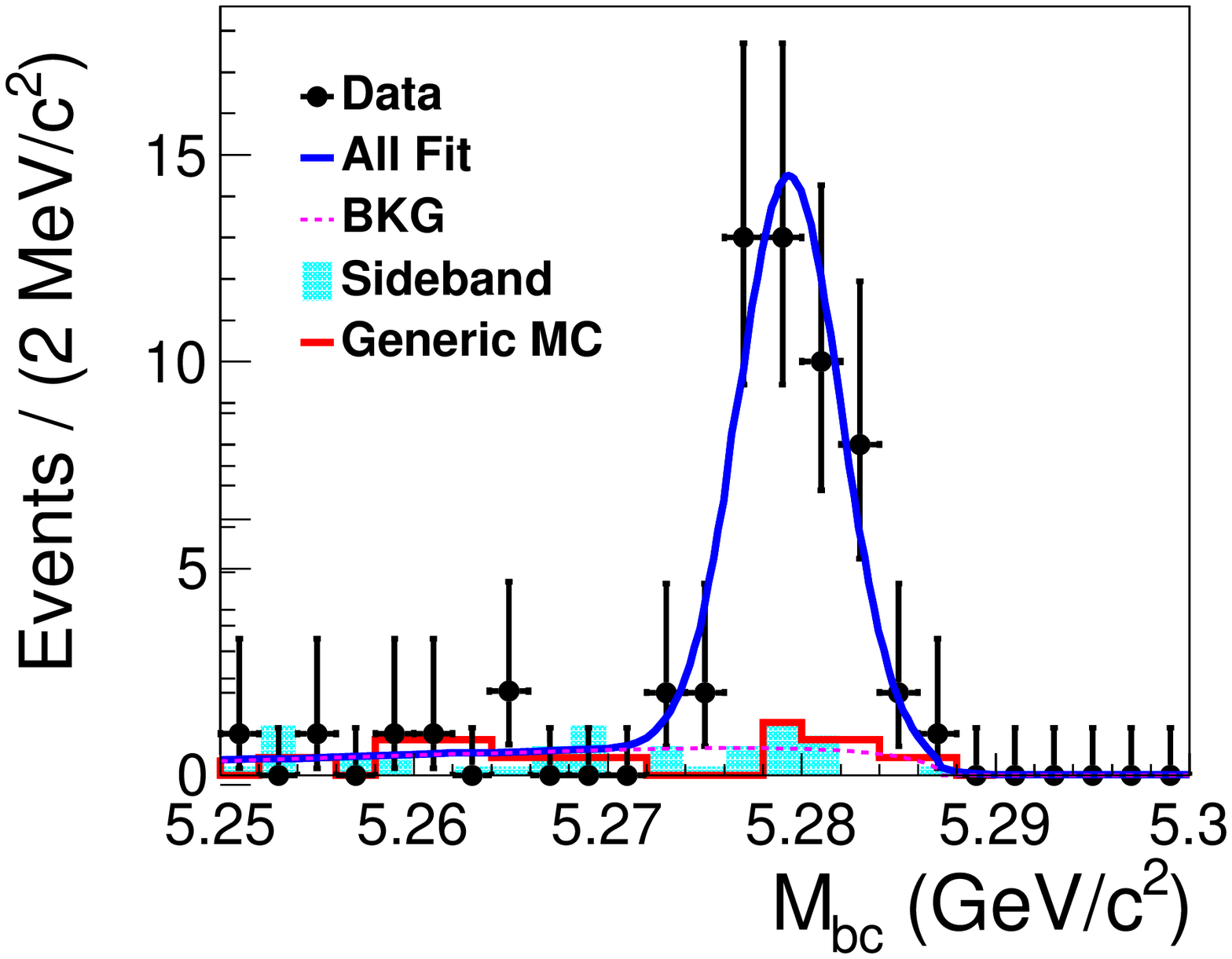}
        \includegraphics[width=5cm]{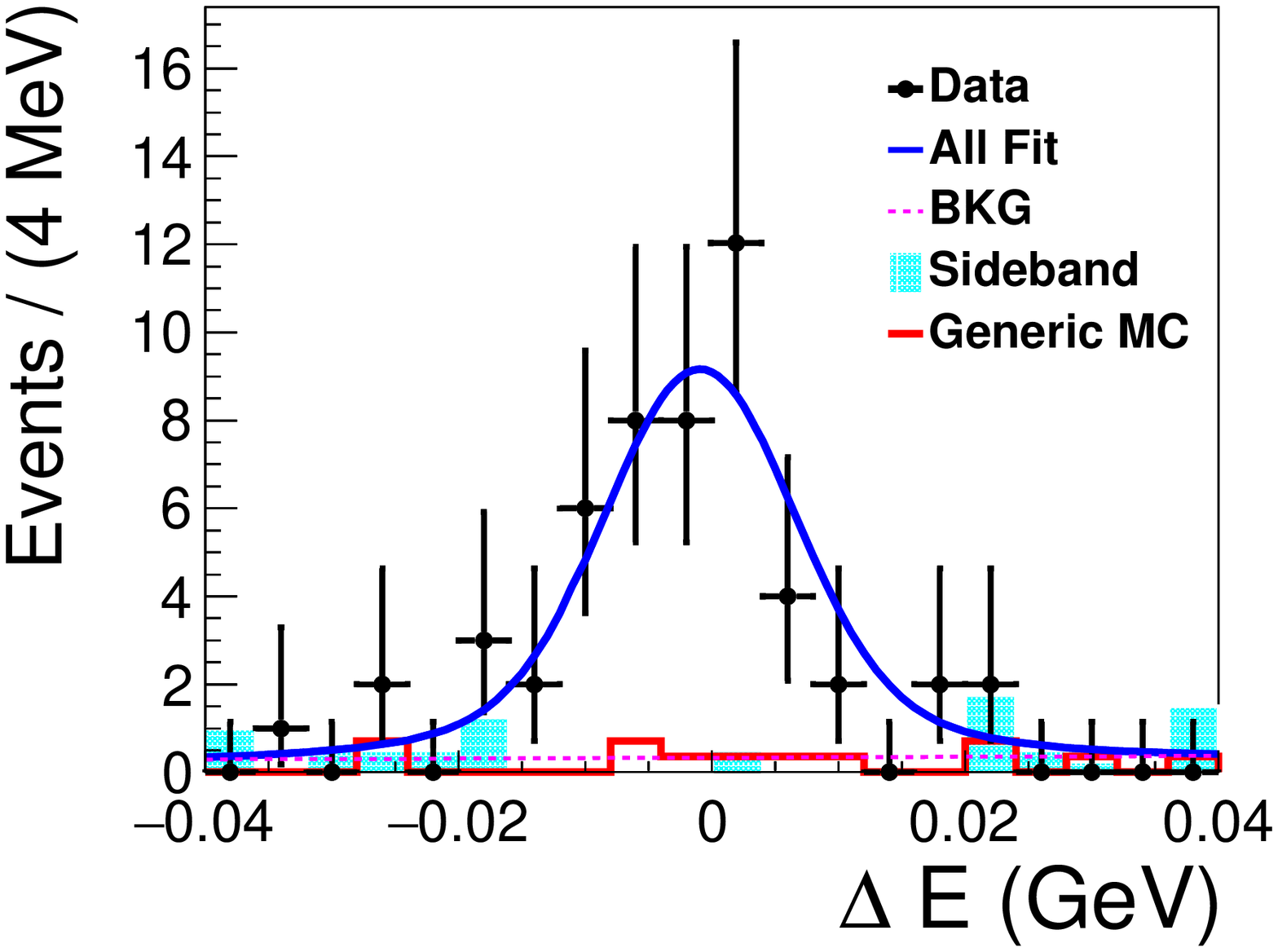}
        \put(-382,79){\bf (a1)} \put(-180,79){\bf (b1)}  \put(-100,79){\bf (c1)}

        \includegraphics[width=5cm,height=3.65cm]{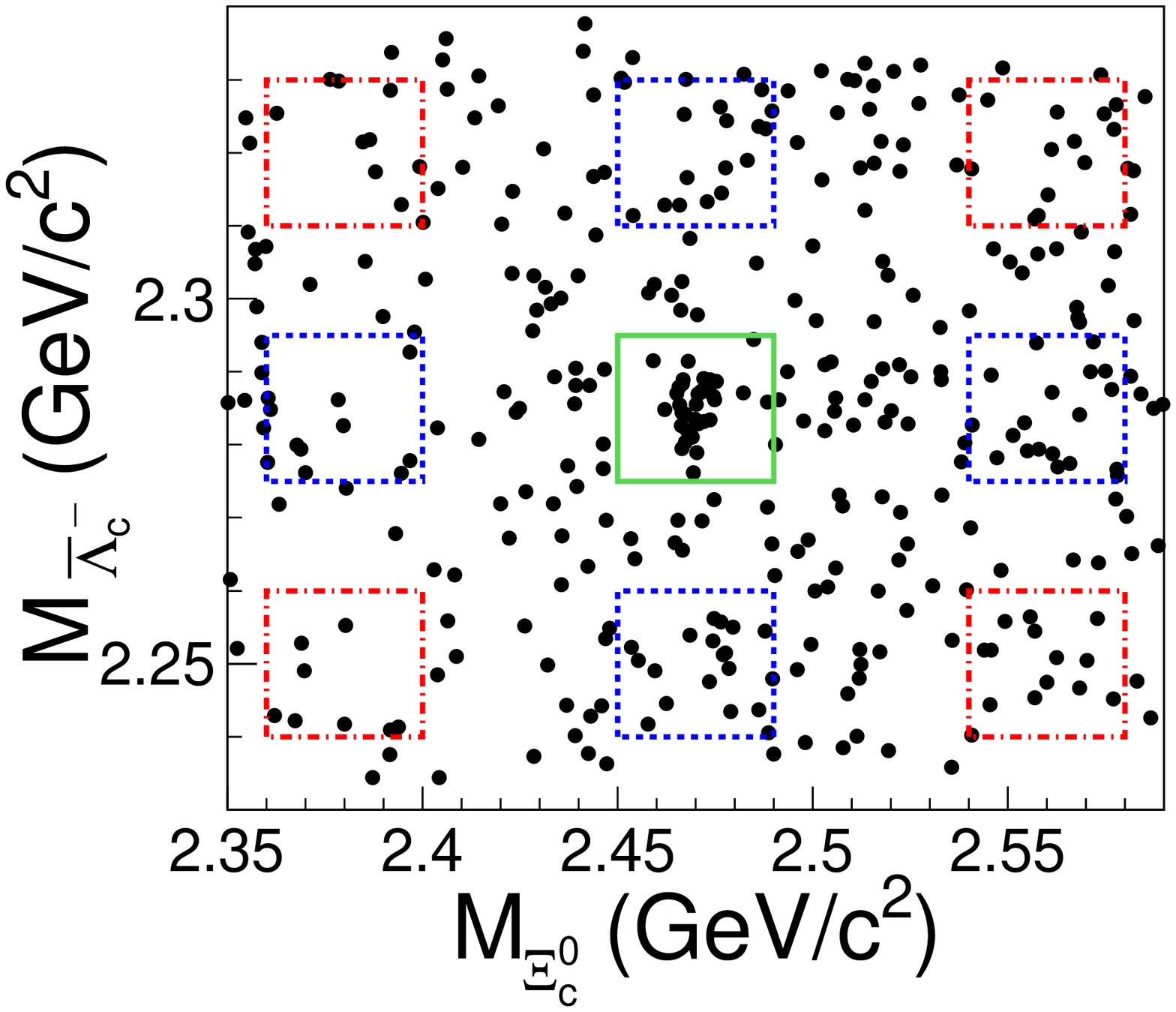}
        \includegraphics[width=5cm]{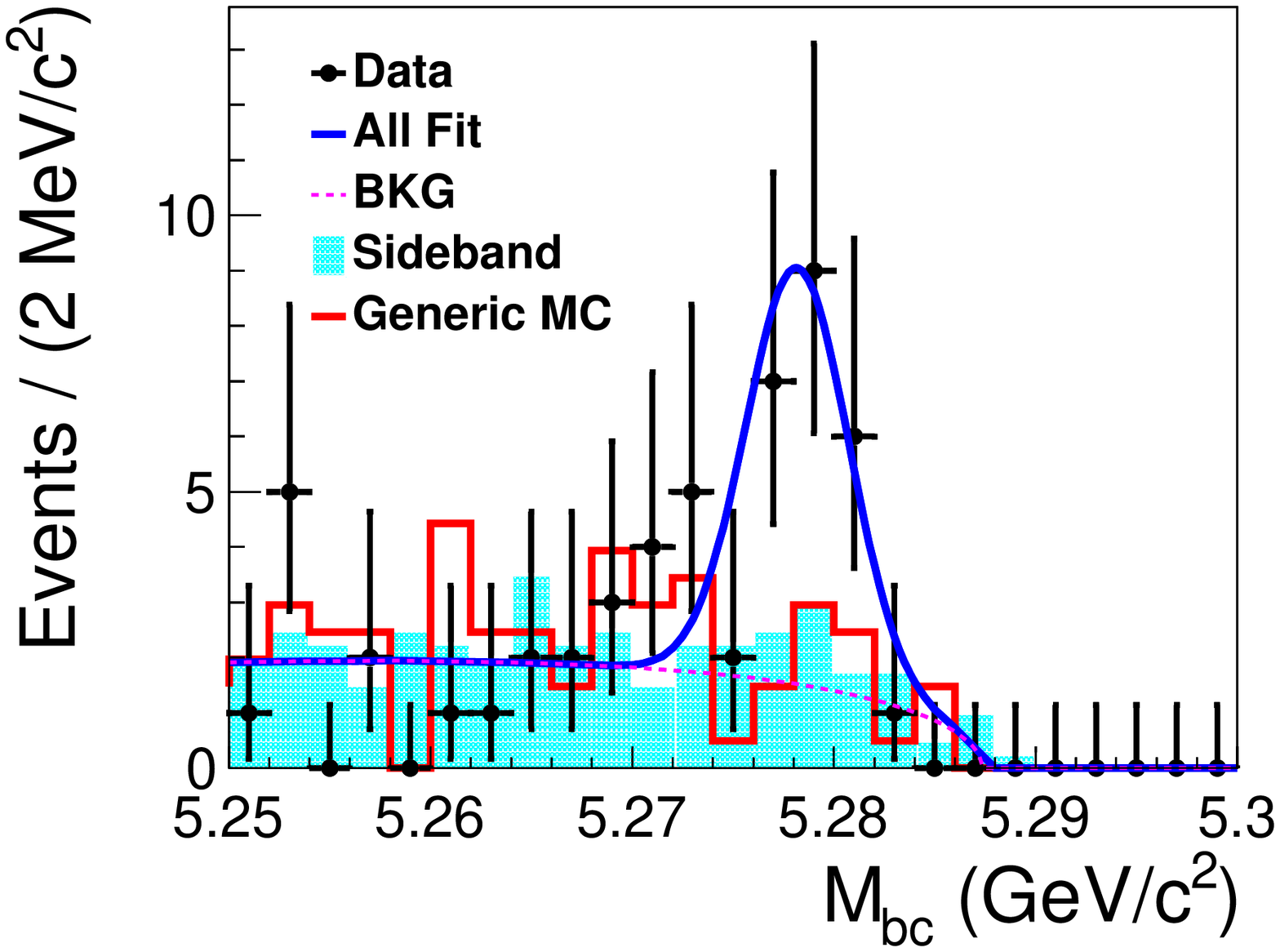}
        \includegraphics[width=5cm]{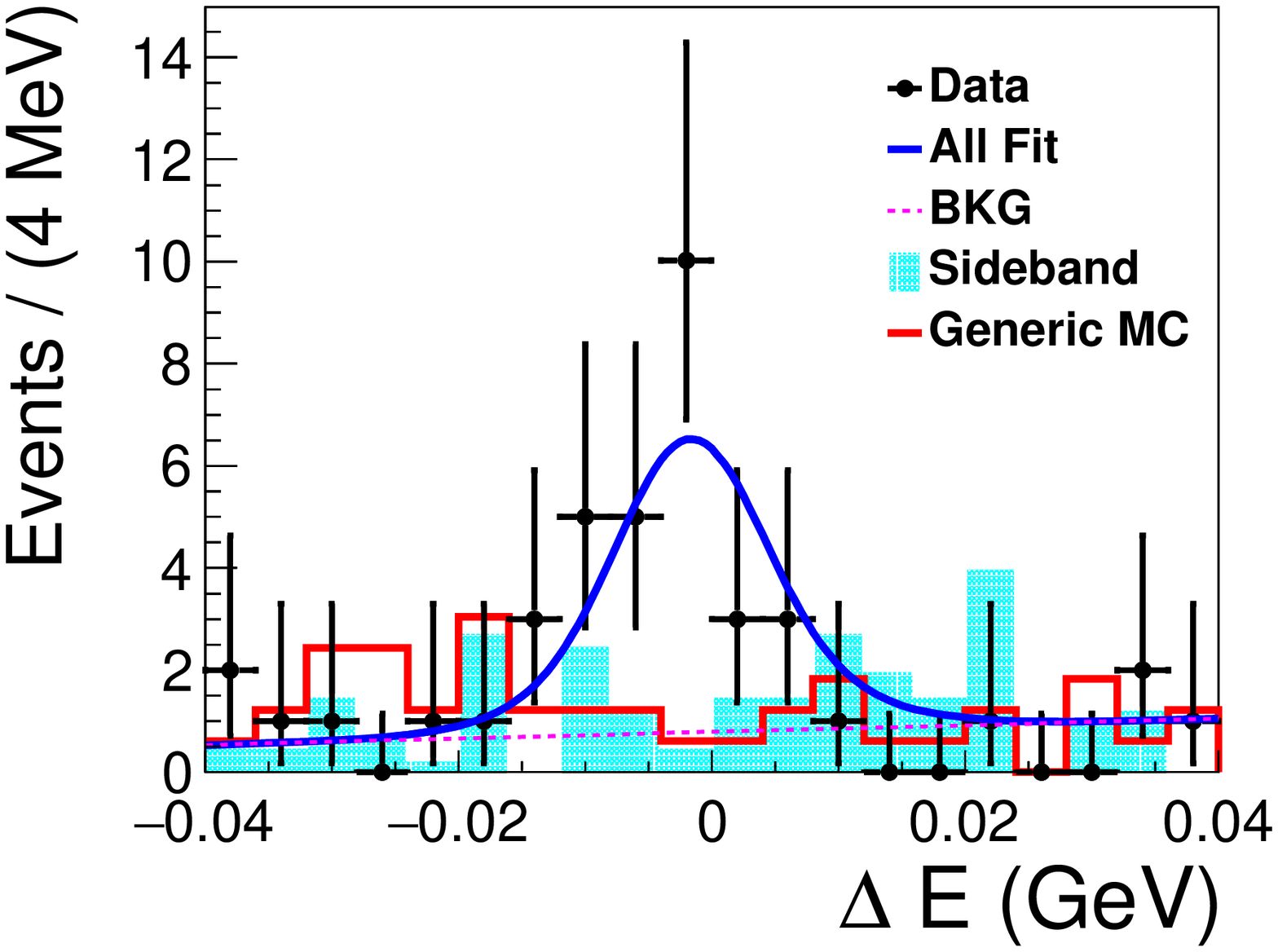}
        \put(-382,79){\bf (a2)} \put(-180,79){\bf (b2)}  \put(-100,79){\bf (c2)}

        \includegraphics[width=5cm,height=3.65cm]{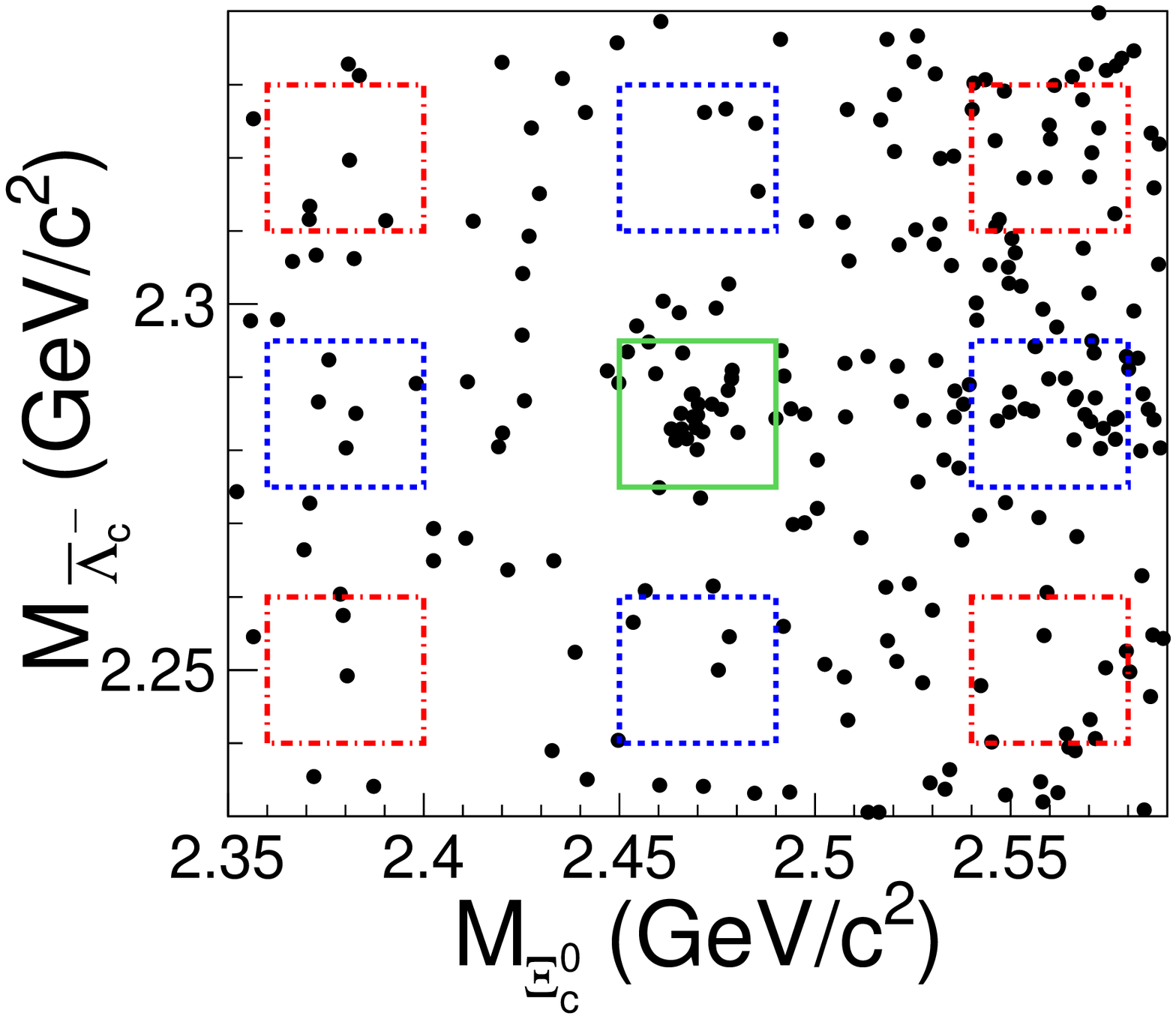}
        \includegraphics[width=4.65cm]{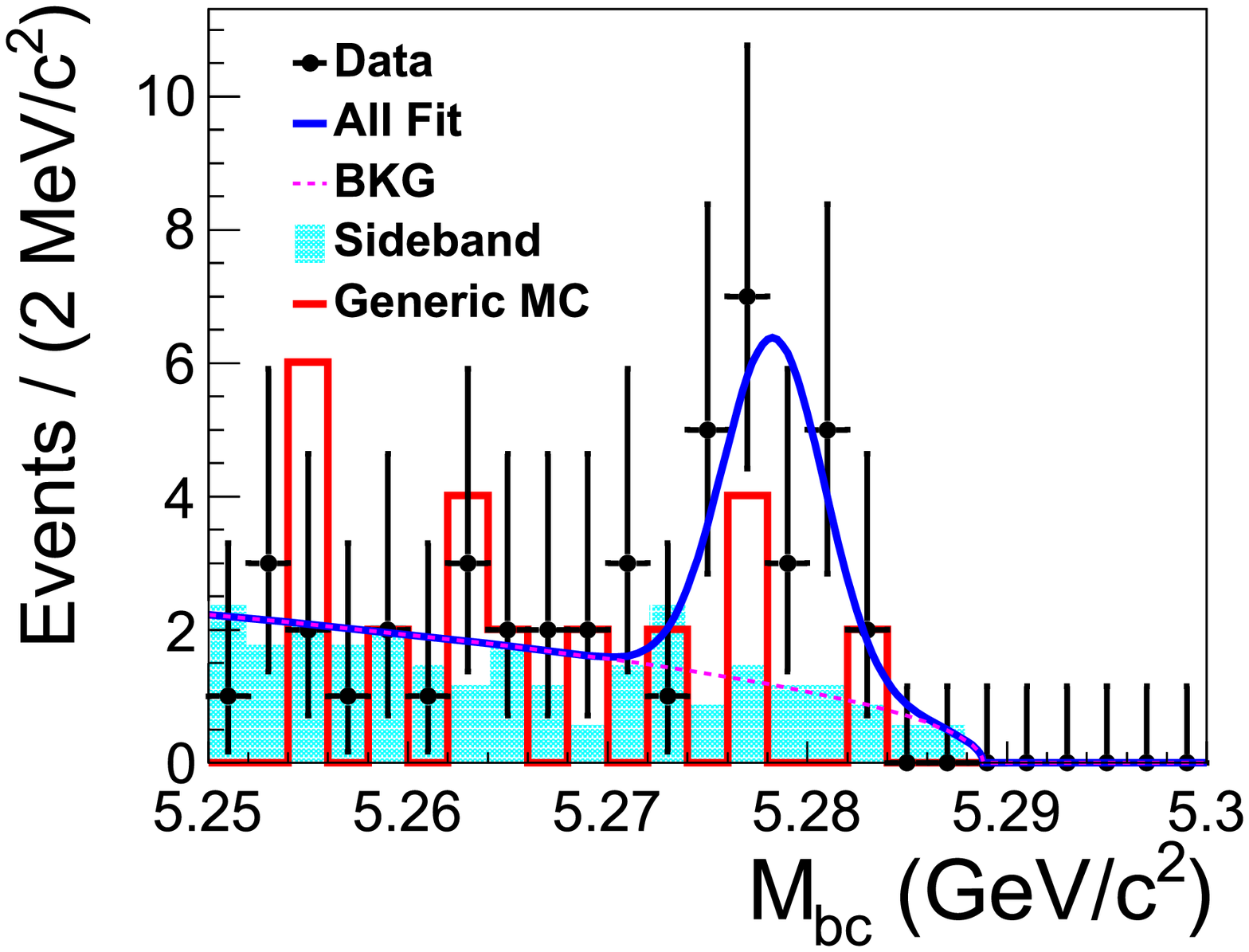}\hspace{0.5cm}
        \includegraphics[width=4.65cm]{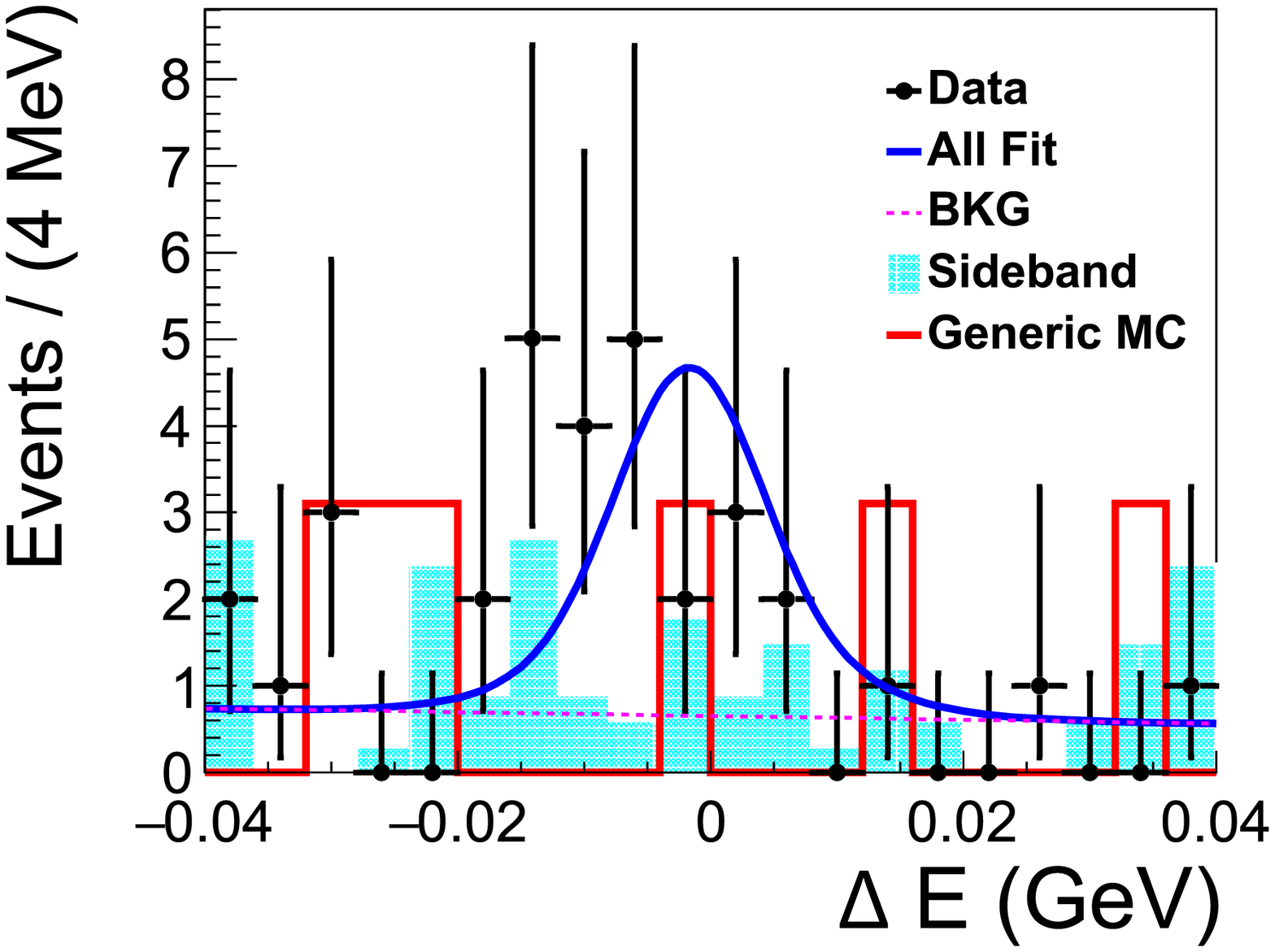}
        \put(-382,79){\bf (a3)} \put(-180,79){\bf (b3)}  \put(-100,79){\bf (c3)}

\caption{The distributions of $M_{\Xi_c^0}$ versus
$M_{\bar{\Lambda}_c^-}$ (a), and the fits to the $M_{\rm bc}$ (b) and
$\Delta E$ (c) distributions of the selected $B^- \to
\bar{\Lambda}_c^- \Xi_c^0$ candidates with $\Xi_c^0 \to \Xi^-
\pi^+$ (1), $\Xi_c^0 \to \Lambda K^- \pi^+$ (2), and $\Xi_c^0 \to
p K^- K^- \pi^+$ (3) decays,
summed over the two reconstructed $\bar{\Lambda}_c^-$
decay modes. In (a), the central solid box defines the signal
region. The red dash-dotted and blue dashed boxes show the
$M_{\Xi_c^0}$ and $M_{\bar{\Lambda}_c^-}$ sideband regions used for
the estimation of the non-$\Xi_c^0$ and non-$\bar{\Lambda}_c^-$
backgrounds (see text). In (b) and (c), the dots with error bars represent the data,
the blue solid curves represent the best fits, and the dashed curves
represent the fitted background contributions. The
shaded and red open histograms have the same meaning as
in Fig.~\ref{FR_xic_com}. } \label{mxic_mlc_data_com}
\end{center}
\end{figure*}

We define a $B^-$ signal region as $M_{\rm bc} > 5.27\,{\rm
GeV}/c^2$ and $|\Delta E| < 0.03\,~{\rm GeV}$. The distributions
of $M_{\Xi_c^0}$ versus $M_{\bar{\Lambda}_c^-}$ for events
in the $B^-$ signal region are shown in Figs.~\ref{mxic_mlc_data_com}(a1) to ~\ref{mxic_mlc_data_com}(a3)
after all selection criteria applied.
The central solid boxes
define the $\Xi_c^0$ and $\bar{\Lambda}_c^-$ signal regions. The
backgrounds from non-$\Xi_c^0$ and non-$\bar{\Lambda}_c^-$ events
are estimated from $M_{\Xi_c^0}$ and $M_{\bar{\Lambda}_c^-}$
sidebands, represented by the dashed boxes in Figs.~\ref{mxic_mlc_data_com}(a1) to ~\ref{mxic_mlc_data_com}(a3).
%The sideband boxes are the same size as the signal box,
The sidebands contribution is estimated
similarly to the inclusive analysis.
Figures~\ref{mxic_mlc_data_com}(b) and \ref{mxic_mlc_data_com}(c) show the
$M_{\rm bc}$ and $\Delta E$ distributions in the $\Xi_c^0$ and
$\bar{\Lambda}_c^-$ signal regions from the selected $B^- \to
\bar{\Lambda}_c^- \Xi_c^0$ candidates with (1) $\Xi_c^0 \to \Xi^-
\pi^+$, (2) $\Xi_c^0 \to \Lambda K^- \pi^+$, and (3) $\Xi_c^0 \to
p K^- K^- \pi^+$. All distributions are summed over the two reconstructed $\bar{\Lambda}_c^-$ decay
modes.

The number of $B^- \to \bar{\Lambda}_c^- \Xi_c^0$ signal events is extracted by performing an unbinned two-dimensional maximum-likelihood fit to the $M_{\rm bc}$ versus $\Delta E$ distributions. For the $M_{\rm bc}$ distribution, the signal shape is modeled with a Gaussian function, and the background is described using an ARGUS function~\cite{argus}. For the $\Delta E$ distribution, the signal shape is modeled using a double-Gaussian function, and the background is described by a first-order polynomial. All shape parameters of the signal functions are fixed to the values obtained from the fits to the MC-simulated signal distributions. The fit results are shown in Fig.~\ref{mxic_mlc_data_com}.

We obtain $N_{\Xi^- \pi^+} = 44.8\pm 7.3$, $N_{\Lambda K^- \pi^+} = 24.1 \pm
5.5 $, and $N_{p K^- K^- \pi^+} = 16.6\pm 5.4$ signal events with statistical
significances of $ 9.5\sigma$, $6.8\sigma$, and $4.6\sigma$.
%for $B^- \to \bar{\Lambda}_c \Xi_c^0$ with $\Xi_c^0 \to \Xi^- \pi^+$,
%$\Lambda K^- \pi^+$, and $p K^- K^- \pi^+$, respectively.
Using the efficiencies calculated from MC simulation, we obtain
 $\BR(B^-\to \bar{\Lambda}_c^- \Xi_c^0) \BR(\Xi_c^0 \to \Xi^- \pi^+) =
       [1.71 \pm 0.28(\rm stat.)] \times 10^{-5},$
 $\BR(B^- \to \bar{\Lambda}_c^- \Xi_c^0) \BR(\Xi_c^0 \to \Lambda K^- \pi^+)
 =  [1.11 \pm 0.26(\rm stat.)] \times 10^{-5},$ and
 $\BR(B^- \to
\bar{\Lambda}_c^- \Xi_c^0) \BR(\Xi_c^0 \to p K^-K^- \pi^+) =
 [5.47 \pm 1.78(\rm stat.)] \times 10^{-6}.$

%%%%%%%%%%%%%%%%%%%%%%%%%%%%%%%%%%%%%
%%%%%%  systematic error  %%%%%%%%%%%
%%%%%%%%%%%%%%%%%%%%%%%%%%%%%%%%%%%%%

There are several sources of systematic uncertainties
as listed in Table~\ref{tab:err}. The reconstruction-efficiency-related
uncertainties include those for tracking efficiency (0.35\% per track),
particle identification efficiency (0.9\% per kaon, 0.9\% per pion, and
3.6\% per proton), as well as $\Lambda$ (3.0\%~\cite{lambda}) and $K_{S}^{0}$
(1.6\%~\cite{kserr}) reconstruction efficiencies. Assuming that all the above
sources of systematic uncertainty are independent, the reconstruction-efficiency-related uncertainties are summed in
quadrature for each decay mode, yielding 4.0--8.4\%, depending
on the specific decay mode. For the four branching-fraction
measurements, the final uncertainties related to the efficiency of the reconstruction are summed in quadrature over
the two reconstructed $\bar{\Lambda}_c^-$ decay modes using weight factors equal
to the product of the total efficiency and the $\bar{\Lambda}_c^-$
partial decay width.

We estimate the systematic uncertainties associated with the fit
by changing the order of the background polynomial, the fitting
range, and by enlarging the mass resolution by 20\%.
The observed deviations are taken as
systematic uncertainties. Uncertainties on $\BR(\bar{\Lambda}_c^- \to
\bar{p} K^+ \pi^-)$ and $\Gamma(\bar{\Lambda}_c^- \to \bar{p}
K_S^0)/\Gamma(\bar{\Lambda}_c^- \to \bar{p} K^+ \pi^-)$ are taken
from Ref.~\cite{PDG}. The final uncertainties on the two $\bar{\Lambda}_c^-$
partial decay widths are summed in quadrature
with the reconstruction efficiency as a weighting factor. The
uncertainty due to the $B$ tagging efficiency is 4.2\%~\cite{FRerr}.
The uncertainty on $\BR[\Upsilon(4S)\to B^+B^-]$ is 1.2\%~\cite{PDG}.
The systematic uncertainty on $N_{\Upsilon(4S)}$ is
1.37\%~\cite{n4serr}.
For the $\Xi_c^0$ branching fractions and the
corresponding ratios, some common systematic uncertainties cancel including
tracking, particle identification,
$\bar{\Lambda}_c^-$ branching fractions, $\Lambda$ and $K_{S}^{0}$ selections,
and $N_{B^-}$.
The sources of uncertainty summarized in Table~\ref{tab:err} are assumed to be independent
and thus are added in quadrature to obtain the total systematic uncertainty.

\begin{table*}[htbp]
\caption{\label{tab:err} Summary of the measured branching fractions and ratios of $\Xi_c^0$ decays (last column), and the corresponding systematic uncertainties (\%). For the branching fractions and ratios, the first uncertainties are statistical and the second are systematic.}
    \begin{tabular}{l||c|c|c|c|c|c||c}
        \hline\hline
        Observable & Efficiency  &  Fit  &  $\Lambda_c$ decays & $B_{\rm tag}$ &\tabincell{c}{ $N_{B^{\pm}}$}  & \tabincell{c}{Sum} & Measured value \\
        \hline
        $\BR(B^- \to \bar{\Lambda}_c^- \Xi_c^0)$ & 3.46 & 4.80  & 5.51 & 4.2 & 1.82 & 9.3 & $(9.51\pm 2.10 \pm 0.88)\times 10^{-4}$ \\
        $\BR( B^- \to \bar{\Lambda}_c^- \Xi_c^0)\BR(\Xi_c^0 \to \Xi^- \pi^+)$ &4.74 &3.49 &5.75  &$...$  & 1.82& 8.4 & $(1.71\pm 0.28 \pm 0.15) \times 10^{-5}$ \\
        $\BR( B^- \to \bar{\Lambda}_c^- \Xi_c^0) \BR(\Xi_c^0 \to \Lambda K^- \pi^+)$  &4.56  & 4.03 &5.82 &$...$   &1.82  &8.6 & $(1.11\pm 0.26\pm0.10)\times 10^{-5}$ \\
        $\BR( B^- \to \bar{\Lambda}_c^- \Xi_c^0) \BR(\Xi_c^0 \to p K^- K^- \pi^+)$  &7.25  & 5.11 &5.03 &$...$   &1.82  &10.5 & $(5.47\pm1.78\pm0.57)\times 10^{-6}$ \\
        $\BR(\Xi_c^0 \to \Xi^- \pi^+)$ & 2.94& 5.9  &$...$ & 4.2 & $...$& 7.8 & $(1.80\pm 0.50 \pm 0.14)\%$ \\
        $\BR(\Xi_c^0 \to \Lambda K^- \pi^+)$ &2.65  & 6.3 & $...$ & 4.2 & $...$  & 8.0 & $(1.17\pm 0.37 \pm 0.09)\%$ \\
        $\BR(\Xi_c^0 \to p K^- K^- \pi^+)$ &3.84  & 7.0 & $...$ & 4.2 & $...$  & 9.0 & $(0.58\pm 0.23 \pm 0.05)\%$  \\ \hline
        $\BR(\Xi_c^0 \to \Lambda K^- \pi^+) /\BR(\Xi_c^0 \to \Xi^- \pi^+)$ &1.36 & 5.3 & $...$ & $...$ & $...$  & 5.5 & $0.65\pm 0.18 \pm 0.04$ \\
        $\BR(\Xi_c^0 \to p K^- K^- \pi^+) /\BR(\Xi_c^0 \to \Xi^- \pi^+)$ &5.24 & 6.2 & $...$ & $...$ & $...$  & 8.1 & $0.32\pm0.12\pm 0.07$ \\
        \hline\hline
    \end{tabular}
\end{table*}

%%%%%%%%%%%%%%%%%%%%%%%%%%%%%%%%%%%%%
%%%%%%%%%%% summary  %%%%%%%%%%%%%%%%
%%%%%%%%%%%%%%%%%%%%%%%%%%%%%%%%%%%%%
In summary, based on $(772 \pm 11) \times 10^{6}$ $B\bar{B}$ pairs collected by Belle,
%collected at the $\Upsilon(4S)$ resonance with the Belle detector at the KEKB asymmetric-energy electron-positron collider,
we have performed an analysis of $B^-
\to \bar{\Lambda}_c^- \Xi_c^0$
inclusively with respect to the $\Xi_c^0$ decay using
a hadronic $B$-tagging method based on a full reconstruction
algorithm~\cite{FR}, and exclusively for $\Xi_c^0$ decays
into $\Xi_c^- \pi^+$, $\Lambda
K^- \pi^+$, and $p K^- K^- \pi^+$ final states.
We report the first measurements of the absolute branching fractions
\begin{displaymath}
\begin{aligned}
\BR(\Xi_c^0 \to \Xi^-\pi^+) & =(1.80 \pm 0.50\pm 0.14)\%, \\
\BR(\Xi_c^0 \to \Lambda K^- \pi^+) &=(1.17 \pm 0.37\pm 0.09)\%,\\
\BR(\Xi_c^0 \to p K^- K^- \pi^+) &=(0.58 \pm 0.23\pm 0.05)\%.
\end{aligned}
\end{displaymath}
The measured $\BR(\Xi_c^0 \to \Xi^-
\pi^+)$ is consistent with the theoretical predictions within uncertainties~\cite{QCD-theory8,Xc-theory2,Xc-theory3}.
The $\BR(B^- \to \bar{\Lambda}_c^- \Xi_c^0)$ is measured for the first time to be
\begin{displaymath}
\begin{aligned}
\BR(B^- \to \bar{\Lambda}_c^- \Xi_c^0)=(9.51 \pm 2.10\pm 0.88) \times 10^{-4}.
\end{aligned}
\end{displaymath}
For the above branching fractions, the first uncertainties are statistical and the second systematic.
The product branching fractions are
$\BR(B^- \to \bar{\Lambda}_c^- \Xi_c^0)\BR(\Xi_c^0 \to \Xi^-
\pi^+)= (1.71 \pm 0.28 \pm 0.15) \times
10^{-5} $, $\BR( B^- \to \bar{\Lambda}_c^- \Xi_c^0) \BR(\Xi_c^0
\to \Lambda K^- \pi^+)= (1.11 \pm 0.26 \pm 0.10) \times 10^{-5}$, and $\BR( B^- \to \bar{\Lambda}_c^- \Xi_c^0) \BR(\Xi_c^0 \to  p K^- K^- \pi^+) = (5.47 \pm 1.78 \pm 0.57) \times 10^{-6}$. The first
two are consistent with previous measurements~\cite{belle-old1,
babar-old2} with improved precision. Our results supersede previous ones from Belle~\cite{belle-old1}.
The ratios of $\BR(\Xi_c^0 \to \Lambda K^- \pi^+)/\BR(\Xi_c^0
\to \Xi^- \pi^+)$ and $\BR(\Xi_c^0 \to p K^- K^-
\pi^+)/\BR(\Xi_c^0 \to \Xi^- \pi^+)$ are $0.65 \pm 0.18 \pm 0.04$ and $0.32 \pm 0.12 \pm 0.07$, respectively, which are consistent
with world-average values $1.07 \pm 0.14$ and $0.34 \pm
0.04$~\cite{PDG} within uncertainties. For the above branching
fractions, the first uncertainties are statistical and the second
systematic.
Our measured $\Xi_c^0$ branching fractions, e.g. that for $\Xi_c^0 \to \Xi^-
\pi^+$, can be combined with $\Xi_c^0$ branching fractions measured relative to
$\Xi_c^0 \to \Xi^- \pi^+$ to yield other absolute $\Xi_c^0$ branching fractions.

%%%%%%%%%%%%%%%%%%%%%%%%%%%%%%%%%%%%%
%%%%%%%   Acknowledgments   %%%%%%%%%
%%%%%%%%%%%%%%%%%%%%%%%%%%%%%%%%%%%%%

We thank Professor Fu-sheng Yu for useful discussions and comments. Y. B. Li acknowledges the support from the China Scholarship Council (201706010043).
We thank the KEKB group for excellent operation of the
accelerator; the KEK cryogenics group for efficient solenoid
operations; and the KEK computer group, the NII, and
PNNL/EMSL for valuable computing and SINET5 network support.
We acknowledge support from MEXT, JSPS and Nagoya's TLPRC (Japan);
ARC (Australia); FWF (Austria); NSFC and CCEPP (China);
MSMT (Czechia); CZF, DFG, EXC153, and VS (Germany);
DST (India); INFN (Italy);
MOE, MSIP, NRF, RSRI, FLRFAS project and GSDC of KISTI and KREONET/GLORIAD (Korea);
MNiSW and NCN (Poland); MSHE, Agreement 14.W03.31.0026 (Russia); ARRS (Slovenia);
IKERBASQUE (Spain);
SNSF (Switzerland); MOE and MOST (Taiwan); and DOE and NSF (USA).


\begin{thebibliography}{99}
\bibitem{charge-conjugate} Inclusion of charge-conjugate states is implicit unless otherwise stated.
\end{thebibliography}
\end{document}